\documentclass[%
 english,
 reprint,
superscriptaddress,
showpacs,
 amsmath,amssymb,
 aps,
prl,
]{revtex4-1}

\usepackage[T1]{fontenc}
\usepackage[latin1]{inputenc}
\usepackage{graphicx}
\usepackage{color}

\usepackage{babel}
\makeatother

\begin{document}

% A useful Journal macro
\def\Journal#1#2#3#4{{#1}\ {\bf #2}, #3 (#4)}

% Some useful journal names
\def\CPL{Chem.\ Phys.\ Lett.}
\def\NCA{Nuovo Cimento}
\def\NIM{Nucl.\ Instrum.\ Methods}
\def\NIMA{Nucl.\ Instrum.\ Methods A}
\def\NPB{Nucl.\ Phys.\ B}
\def\PLB{Phys.\ Lett.\ B}
\def\PRL{Phys.\ Rev.\ Lett.}
\def\PRB{Phys.\ Rev.\ B}
\def\PRD{Phys.\ Rev.\ D}
\def\ADNDT{At.\ Data Nucl.\ Data Tabl.}
\def\ZPB{Z.\ Phys.\ B}
\def\ZPC{Z.\ Phys.\ C}
\def\SSC{Solid State Commun.}
\def\SSCH{Solid State Chem.}
\def\JES{J.\ Electr. Spectr.\ Relat.\ Phen.}
\def\NAT{Nature}
\def\AC{Acta Cryst.}
\def\ACA{Acta Cryst.\ A}
\def\ACB{Acta Cryst.\ B}
\def\PB{Physica B}
\def\PC{Physica C}
\def\LCM{Less.-Common Met.}
\def\EPL{Europhys.\ Lett.}
\def\JS{J.\ Super.}
\def\JCG{J.\ Cryst.\ Growth}
\def\JAC{J.\ Appl.\ Crystallogr.}
\def\JPSJ{J.\ Phys.\ Soc.\ Jpn.}
\def\JACS{J.\ Am.\ Chem.\ Soc.}
\def\JLTP{J.\ Low Temp.\ Phys.}
\def\JPCM{J.\ Phys.\ Cond.\ Matt.}
%\Journal{\PRB}{52}{13911}{1995}

%%\renewcommand{\textfraction}{0.02}

%%\preprint{Merz-2011/Sr(Fe$_{1-x}$Co$_x$)$_2$As$_2$  \today}

\title{Electronic structure of single-crystalline Sr(Fe$_{1-x}$Co$_x$)$_2$As$_2$ probed by x-ray absorption spectroscopy: Evidence for isovalent substitution of Fe$^{2+}$ by Co$^{2+}$}

\author{M.\@ Merz}
\email[Corresponding author. ]{michael.merz@kit.edu}
\affiliation{Institut f\"{u}r Festkörperphysik, Karlsruhe Institute of Technology, 76021 Karlsruhe, Germany}

\author{F.\@ Eilers}
\affiliation{Institut f\"{u}r Festkörperphysik, Karlsruhe Institute of Technology, 76021 Karlsruhe, Germany}
\affiliation{Fakult\"{a}t für Physik, Karlsruhe Institute of Technology, 76031 Karlsruhe, Germany}

\author{Th.\@ Wolf}
\affiliation{Institut f\"{u}r Festkörperphysik, Karlsruhe Institute of Technology, 76021 Karlsruhe, Germany}

\author{P.\@ Nagel}
\affiliation{Institut f\"{u}r Festkörperphysik, Karlsruhe Institute of Technology, 76021 Karlsruhe, Germany}

\author{H.\@ v.\@ L\"{o}hneysen}
\affiliation{Institut f\"{u}r Festkörperphysik, Karlsruhe Institute of Technology, 76021 Karlsruhe, Germany}
\affiliation{Physikalisches Institut, Karlsruhe Institute of Technology, 76131 Karlsruhe, Germany}

\author{S.\@ Schuppler}
\affiliation{Institut f\"{u}r Festkörperphysik, Karlsruhe Institute of Technology, 76021 Karlsruhe, Germany}

\date{\today}

\begin{abstract}
The substitutional dependence of valence and spin-state configurations of Sr(Fe$_{1-x}$Co$_x$)$_2$As$_2$ ($x =$ 0, 0.05, 0.11, 0.17, and 0.38) is investigated with near-edge x-ray absorption fine structure at the $L_{2,3}$ edges of Fe, Co, and As. The present data provide direct spectroscopic evidence for an effectively isovalent substitution of Fe$^{2+}$ by Co$^{2+}$, which is in contrast to the widely assumed Co-induced electron-doping effect.  Moreover, the data reveal that not only does the Fe valency remain completely unaffected across the entire doping range, but so do the Co and As valencies as well. The data underline a prominent role of the hybridization between (Fe,Co) 3$d_{xy}$, $d_{xz}$, $d_{yz}$ orbitals and As $4s/4p$ states for the band structure in $A$(Fe$_{1-x}$Co$_x$)$_2$As$_2$ and suggest that the covalency of the (Fe,Co)-As bond is a key parameter for the interplay between magnetism and superconductivity.

\end{abstract}


%\keywords{Suggested keywords}%Use showkeys class option if keyword
                              %display desired
\maketitle

%%\section{Introduction}

Since the discovery of superconductivity in iron-based materials~[\onlinecite{Kamihara2008}] with transition temperatures $T_c$ up to 55~K~[\onlinecite{Wang2008}], the iron pnictides and chalcogenides have been intensely studied, and the understanding of the physical properties of these systems has considerably advanced. As for other materials like heavy-fermion systems and high-$T_c$ cuprates, superconductivity emerges in the vicinity of a magnetic instability. Moreover, similar to the cuprates, the superconducting phase boundary of most iron pnictides has a dome-like shape and superconductivity appears when the antiferromagnetic (AFM) phase is significantly reduced upon either substitution on the transition-metal and on the pnictide site or upon external pressure. Even though it has been established that the edge-sharing FeAs$_4$ tetrahedra are the structural key ingredient for the physical properties of the pnictides, the details of the superconducting pairing mechanism remain still elusive. Meanwhile, however, many studies strongly suggest that distinct nesting properties are important for the magnetic as well as for the superconducting characteristics.\\[0.25mm]
\indent For the stoichiometric parent compound, the good nesting properties, which are closely connected with interband scattering between Fermi-surface hole pockets at the $\Gamma$ point of the Brillouin zone and electron pockets at the $M$ point, are obviously responsible for the development of a spin density wave (SDW) and AFM order at low temperature \cite{Yin2008,Singh2008,Singh2008b,Singh2009,Mazin2008,Mazin2009}. Upon doping (or application of pressure) the shape and the size of the hole and electron pockets are modified \cite{Yin2008,Mazin2008,Kuroki2008}, the SDW is suppressed, and specific spin fluctuations associated with \textit{close-to-nesting} conditions are assumed to play a decisive role for the superconducting pairing mechanism. Whether magnetism and superconductivity do indeed coexist 
\begin{figure*}
\hspace{-2.5mm}
\includegraphics[width=0.7\textwidth]{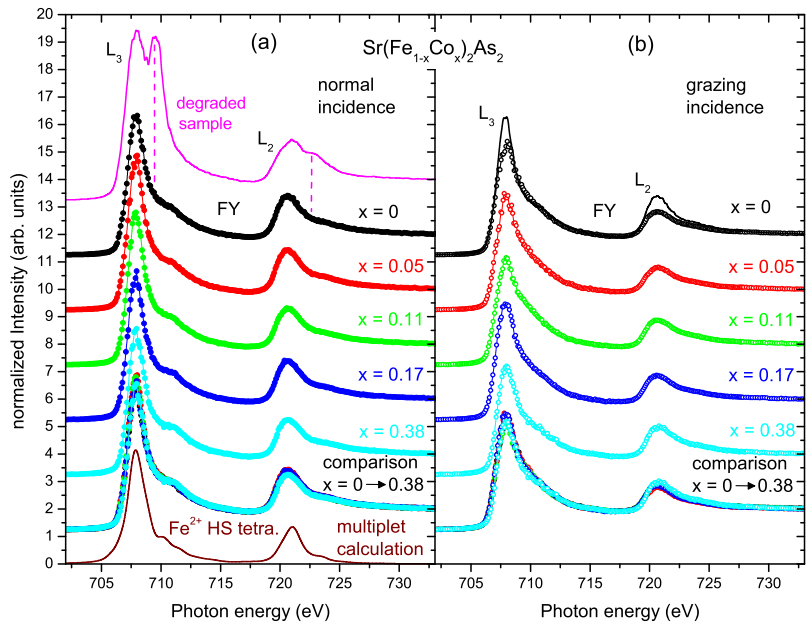}

\caption{\label{fig1}(Color online) Comparison of the (a) normal and (b) grazing incidence Fe $L_{2,3}$ NEXAFS spectra of Sr(Fe$_{1-x}$Co$_x$)$_2$As$_2$ ($x =$ 0, 0.05, 0.11, 0.17, and 0.38) recorded at 300 K in FY. The spectral shape of both edges is unaffected upon doping. For clarity, the spectra are vertically offset. In addition, the spectrum of a deteriorated iron-oxide containing sample is included in (a) (topmost panel), thereby exhibiting the respective peak positions of the iron oxide. For more direct comparison of the doping dependence, the spectra are plotted on top of each other in the next to lowest panel. The multiplet calculations show that the spectra can be described reasonably well for tetrahedrally coordinated Fe$^{2+}$\@. As a representative, in the topmost panel in (b) the anisotropy between normal (line) and grazing incidence (symbols) is depicted for the $x= 0$ sample.}
\end{figure*}
in the pnictides \cite{Marsik2010,Long2011} is still under intense debate, as is the related question whether the same electrons establish AFM order and are responsible for superconductivity \cite{Singh2009}. Furthermore, the band structure is very complicated as well in these systems since it results from a compromise between the covalent Fe-As bonds (for the tetrahedral coordination the $d_{xy}$, $d_{xz}$, and $d_{yz}$ states are higher in energy than the $d_{x^2-y^2}$ and the $d_{3z^2-r^2}$ states) and the square planar Fe-Fe coordination (where the energy sequence descends as $d_{x^2-y^2}-d_{xy}-d_{xz},d_{yz}-d_{3z^2-r^2}$ in an ionic picture). 
As a consequence, all bands with Fe $3d$ character cross the Fermi energy, thereby reflecting the itinerant character of the system.\\[0.25mm]
\indent An interesting example is the $A$Fe$_2$As$_2$ system (where $A$ denotes Ca, Sr, or Ba) which exhibits two-band superconductivity for the following cases: (i) an external pressure of several GPa is applied to the samples \cite{Alireza2009,Yamazaki2010,Drotziger2010}, (ii) As is partially substituted by isovalent P \cite{Jiang2009,Kasahara2010}, (iii) Fe is partially replaced by isovalent Ru \cite{Sharma2010,Rullier2010,Thaler2010}, (iv) hole doping achieved by partial replacement of $A^{2+}$ by K$^{+}$ \cite{Rotter2008b,Rotter2008c}, and (v) Fe is partially substituted by Co, Ni, Co/Cu mixtures, Rh, Pd, Ir, or Pt \cite{Canfield2010,Ni2009,Kim2011,Wang2010,Zhu2010,Saha2010}. Although the valencies of Co, Ni, Cu, Rh, Pd, Ir and Pt in $A$Fe$_2$As$_2$ are still unclear, ``electron doping'' is generally assumed for the substitutions under~(v). In the framework of the so-called virtual crystal approach, a
rigid-band shift of $E_F$ induced by ``electron doping'' is considered \cite{Sefat2008}. Density-functional theory (DFT) shows that the main effect of substitution is not on the density of states but rather on the
\begin{figure*}[t]
\hspace{-2.5mm}
\includegraphics[width=0.72\textwidth]{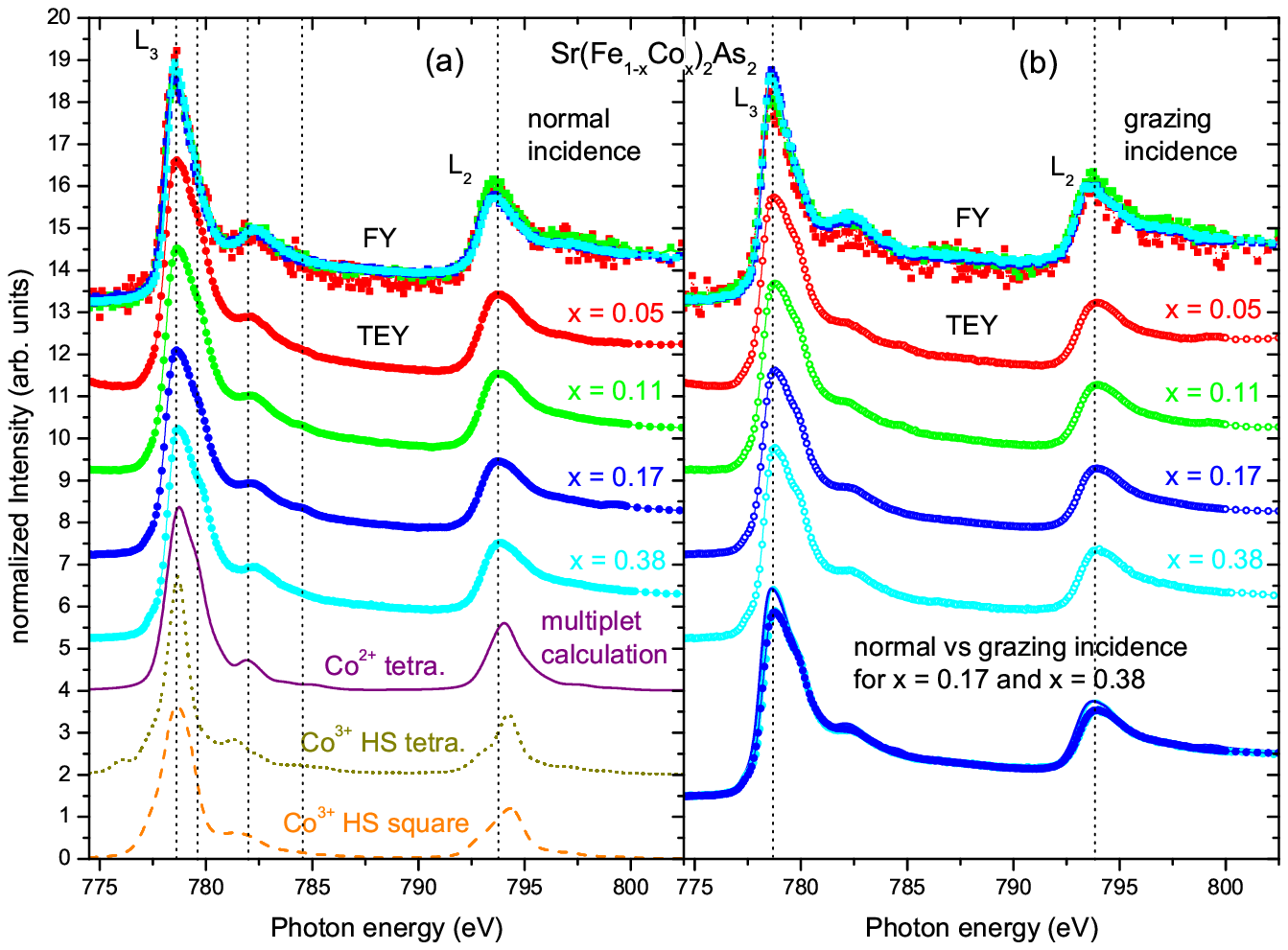}

\caption{\label{fig2}(Color online) Comparison of the (a) normal and (b) grazing incidence Co $L_{2,3}$ NEXAFS spectra of Sr(Fe$_{1-x}$Co$_x$)$_2$As$_2$ ($x =$ 0, 0.05, 0.11, 0.17, and 0.38) taken at 300 K in FY (topmost panel) and in TEY. The spectral shape of both edges is unaffected upon doping. For clarity, the TEY spectra are vertically offset. The multiplet calculations show that the spectrum can be described reasonably well for tetrahedrally coordinated Co$^{2+}$\@. As a representative, in the lowest panel in (b) the anisotropy between normal (dotted line) and grazing incidence (symbols) is depicted for the $x= 0.17$ and the $x= 0.38$ sample.}
\end{figure*}
relative size of the electron and hole pockets \cite{Singh2008b}. More recent DFT calculations even indicate that Co does not dope the system with electrons at all but is isovalent to Fe$^{2+}$ and solely acts as a random scatterer \cite{Wadati2010}. In this case, the effect of Co substitution would rather lead to a topological change of the Fermi surface which destabilizes magnetism in favor of superconductivity.\\[0.25mm]
\indent To shed more light on the substitution-dependent changes in the electronic structure of the iron pnictides and to scrutinize the supposed ``electron
doping'' by analyzing the valency of the relevant chemical elements, we have investigated the Sr(Fe$_{1-x}$Co$_x$)$_2$As$_2$ system ($x =$ 0, 0.05, 0.11, 0.17, and 0.38) \cite{footnote1} with near-edge x-ray absorption fine structure (NEXAFS) at the $L_{2,3}$ edges of
Fe, Co, and As using linearly polarized light. Since it has been shown that iron pnictide samples may suffer severely from iron oxide contamination \cite{Parks2010}, the O $K$ edge was investigated as well. For completeness the results obtained at the O $K$ edges are shown and
discussed in the supplementary material. The measurements were performed
at the Institut für Festkörperphysik beamline WERA
at the ANKA synchrotron light source (Karlsruhe, Germany). All spectra were taken simultaneously in bulk-sensitive fluorescence yield (FY)\@ and in total electron yield (TEY) on single crystals that exhibit a mirror-like shiny surface. Photon energy calibration to better than 30 meV was ensured by adjusting the Ni $L{}_{3}$ peak position measured on a NiO single crystal before and after each NEXAFS scan to the established peak position \cite{Merz2010}. The spectral resolution was set to 0.3~eV for the Fe and Co $L_{2,3}$ edges. 
While the in-plane spectrum is obtained for a normal-incidence alignment, i.~e.\@, for a grazing angle $\theta$\@ of 0$^{\circ}$\@, the out-of-plane spectrum is determined by measuring in a grazing-incidence setup with a grazing angle of 65$^{\circ}$\@. The FY spectra are corrected for the self-absorption and saturation effects inherent to the FY method. Utilizing multiplet calculations, the configuration of the corresponding spin and valence states of Fe and Co is determined for the investigated doping contents \cite{commentMerz31}. Sr(Fe$_{1-x}$Co$_x$)$_2$As$_2$ single crystals were grown from self-flux in glassy carbon crucibles as described elsewhere \cite{Hardy2009,Hardy2010}. The composition of the samples was determined using energy dispersive x-ray spectroscopy and was verified by the size of the respective background-corrected edge jump in our NEXAFS experiments. Using superconducting quantum interference device (SQUID) magnetometry, the superconducting transition temperature was determined to $T_c \approx 12$~K for the nearly optimally doped $x=0.11$ sample, to $\approx 9$~K for the slightly overdoped $x=0.17$ sample, and to $\approx 7$~K for the strongly overdoped $x=0.38$ sample. No superconducting transition down to 4 K was observed for the undoped $x=0$ and the underdoped $x=0.05$ sample. Thus, the studied samples span the entire range from undoped up to highly overdoped.\\[0.25mm]
\indent In Fig\@.~\ref{fig1} the Fe $L_{2,3}$ NEXAFS spectra of Sr(Fe$_{1-x}$Co$_x$)$_2$As$_2$ ($x =$ 0, 0.05, 0.11, 0.17, and 0.38) measured at 300 K in FY are depicted for (a) normal and (b) grazing incidence. The
absorption spectra correspond to first order to transitions of the type Fe~2\emph{$p^{6}$}3$d^{6}$ $\rightarrow$ Fe 2$p^{5}$3$d^{7}$\@. They consist of two manifolds of multiplets located around 708~eV~($L_{3}$) and 721~eV~($L_{2}$) and separated by the spin-orbit splitting of the Fe 2\emph{p} core level. In addition, the spectrum of a degraded undoped sample is included in Fig\@.~\ref{fig1}~(a) (topmost panel), exhibiting the respective additional peak positions of iron oxide around $\approx 710$ and $\approx 723$~eV. It is obvious from the absence of these two oxide-related features in all other spectra in Fig\@.~\ref{fig1} that iron oxide does not play an important role for the bulk properties of the investigated series. [The TEY data of most samples (not shown) are similar to the FY spectrum of the degraded sample which reflects that the surface layer contains some iron oxide phase as well (see also discussion of O $K$ edge in the supplementary material).]\\[0.25mm]
\indent It is also evident from the FY data that the spectral shape of both edges remains unaffected upon Fe substitution by Co, as do the respective energetic position of the onset energy of the $L_{3}$ and the $L_{2}$ edge and of all other spectral features (like peaks or shoulders). This is an important finding since it establishes that the Fe valency is unchanged across the entire doping series. It should be noted that for electron doping, changes in the spectral shape are expected together with a spectral shift $\Delta E$ to lower energies which according to multiplet calculations scales in this energy range almost linearly with the doping content $x$ as $\Delta E \approx x \times 1.5$~eV.
If the assumption of electron doping were valid \cite{commentBittar,Bittar2011}, and with the good energy precision in our experiment, it should be possible to observe - at least for the highly doped $x =$ 0.17 and 0.38 samples - a modified spectral shape and a significant energy shift (of $\approx$ 0.26 and 0.57 eV, resp.). Such a shift was indeed found for NEXAFS on electron-doped LaFeAsO$_{1-x}$F$_x$ \cite{Kroll2008} - but does not exist here in the Sr(Fe$_{1-x}$Co$_x$)$_2$As$_2$ system with partial replacement of Fe by Co. Furthermore, our multiplet calculations show that the spectrum can be described reasonably well for tetrahedrally coordinated Fe$^{2+}$\@ in a high-spin (HS) configuration. To obtain the simulated data, the code developed by Thole, Butler, and Cowan \cite{Thole1987,Butler1962,Cowan1981} and maintained and further developed by de Groot \cite{Stavitski2010} was used to calculate spectra for different values of the crystal-field splitting $\Delta_{\rm CF}$ and of the charge-transfer energy $\Delta_{c}$\@, and by taking the Hund's rule exchange interaction into account. Charge-transfer effects were included for Fe$^{2+}$ by admixing transitions of the type 2\emph{$p^{6}$}3$d^{7}\underline{L}$ $\rightarrow$ 2$p^{5}$3$d^{8}\underline{L}$, where $\underline{L}$ denotes a hole at the arsenic ligand \cite{footnote3}. Consistent with the TEY spectra in Ref.~\cite{Parks2010} a certain anisotropy between normal and grazing incidence data can, independent of the Co content, be inferred from Fig\@.~\ref{fig1} with the higher density of states found for the in-plane spectra.\\[0.25mm]
\indent In Fig\@.~\ref{fig2}~(a) and (b) the normal- and grazing-incidence Co $L_{2,3}$ NEXAFS spectra of Sr(Fe$_{1-x}$Co$_x$)$_2$As$_2$ ($x =$ 0.05, 0.11, 0.17 and 0.38) recorded at 300 K in FY and in TEY are displayed. As for the Fe $L_{2,3}$ edge, they consist of two manifolds of multiplets, located around 778~eV~($L_{3}$) and 794~eV~($L_{2}$). The comparison between the bulk-sensitive FY [topmost panel in figure (a) and (b)] and the TEY data shows a close resemblance between the respective spectra implying the absence of a cobalt-oxide surface layer. Due to the much better statistics of the TEY data we, therefore, 
focus on the TEY spectra in the following. Similarly to the Fe edge (see above), no 
concentration-dependent spectral changes or energetic shifts are observed at the Co $L_{2,3}$ edge for the $x =$ 0.05, 0.11, 0.17, and 0.38 samples.
Again a certain anisotropy between normal and grazing incidence data can be attributed to the higher density of states found
for the in-plane spectra. The fact that the spectral shape of the Co $L_{2,3}$ spectra is completely independent of $x$, as is the energetic position of the onset energy of the $L_{3}$ and the $L_{2}$ edge and of all other spectral features, unambiguously demonstrates that the Co valency remains the same throughout the entire investigated doping series.\\[0.25mm]
\indent To derive a reliable estimate of the Co valency and the corresponding spin state, we have performed multiplet calculations for Co$^{2+}$ and Co$^{3+}$ in tetrahedral and square-planar coordination, respectively \cite{footnote3}. The spectral shape of square-planar coordinated Co$^{2+}$ HS and Co$^{3+}$ low-spin (LS) (not shown) does not show any resemblance to our experimental data. Hence, this configuration can be excluded right away. Furthermore, a Co$^{3+}$ LS in tetrahedral coordination is stabilized only for an unusually strong crystal-field splitting of $\Delta_{\rm CF} \gtrsim 3.5$~eV\@.  Therefore, only the simulated spectra of Co$^{2+}$ and Co$^{3+}$ HS in tetrahedral coordination and Co$^{3+}$ HS in square-planar coordination are considered in Fig\@.~\ref{fig2}\@. By aligning the calculated spectra to the energy position of the main structure of the measured $L_3$ peak position, it is evident that only Co$^{2+}$ in tetrahedral coordination (solid line) reproduces reasonably well the measured spectra for $x =$ 0.05, 0.11, 0.17, and 0.38. Even the shoulder at the $L_3$ edge around 779.5 eV and the small wiggles above 780 eV are adequately described for Co$^{2+}$ in tetrahedral coordination. For Co$^{3+}$ HS in tetrahedral (dotted line) or square-planar coordination (dashed line), additional satellites appear in the calculated spectrum at $\approx 776$, 777, and 781 eV. Furthermore, both the energy position of the small peak around 782 eV and of the complete $L_2$ edge cannot be reconciled with the experimental data \cite{footnote2}. 
It is interesting to note that already the experimental Fe $L_{2,3}$ spectra in Fig\@.~\ref{fig1} are best described by means of a tetrahedrally coordinated Fe atom. It should also be mentioned that, in order to properly describe the experimental data, the crystal-field parameter of the multiplet calculations increases from $\Delta_{\rm CF} \approx 1.0$~eV for FeAs$_4$ to $\approx 1.4$~eV for CoAs$_4$\@ while the charge-transfer energy decreases systematically from $\Delta_c \approx 3.0$~eV to $\approx 1.0$~eV \cite{footnote3}. Both effects point to a more pronounced covalent character of the Co-As bond.\\[0.25mm]
\indent To complement the electronic structure derived from the Fe and Co $L_{2,3}$ edges, 
\begin{figure}[b]
\hspace{-3.5mm}
\includegraphics[width=0.5\textwidth]{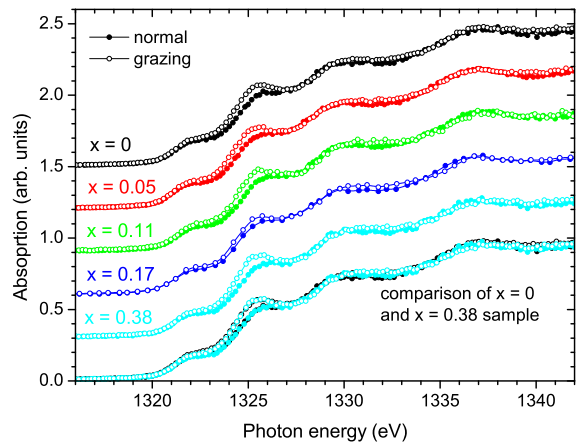}

\caption{\label{fig3}(Color online) Comparison of the normal and grazing incidence As $L_{3}$ NEXAFS spectra of Sr(Fe$_{1-x}$Co$_x$)$_2$As$_2$ ($x =$ 0, 0.05, 0.11, 0.17 and 0.38) recorded at 300 K in FY. The spectral shape of both edges is almost unaffected upon doping. For clarity, the spectra are vertically offset. A small anisotropy appears for all doping contents between normal and grazing incidence spectra. For more direct comparison of the doping dependence, the spectra of the $x =$ 0 and $x =$ 0.38 sample are plotted on top of each other in the lowest panel.}
\end{figure}
the normal- and grazing-incidence As~$L_{3}$ NEXAFS spectra of Sr(Fe$_{1-x}$Co$_x$)$_2$As$_2$ ($x =$ 0, 0.05, 0.11, 0.17 and 0.38) taken at 300 K with an energy resolution of 1.1~eV are depicted in Fig\@.~\ref{fig3}. These spectra are very similar to the data given in
Ref.~\cite{Parks2010}. Although probing the As~$L_{2,3}$ edge is only an indirect measure of the $4s/4p$ states, it was shown in Ref.~\cite{Parks2010}
that the polarization dependence of the As~$L_{3}$ spectra allows the
determination of the As orbital topology. 
Consistent with the previous data, the present spectra point to a small anisotropy between in-plane and out-of-plane states with a slightly higher density of states found for the out-of-plane spectra. More important is, however, that significant changes in the spectral shape and/or energetic shifts are not observed in the spectra upon substitution of Fe by Co up to almost 40\%. This indicates that the As valency remains completely unchanged upon doping as well. Hence neither at the As nor at the Fe atom the ``extra electron'' expected for Co$^{3+}$ substitution is found - lending further support to the notion that Co, instead of behaving like Co$^{3+}$, seems to appear here with a valency of $+2$. Recent calculations \cite{Wadati2010,Levy2012,Berlijn2012,Liu2012} seem to agree, and suggest short-range screening as an important factor: in this picture, the ensemble of $d$ electrons in Sr(Fe$_{1-x}$Co$_x$)$_2$As$_2$ screens the extra positive charge of Co$^{3+}$ on such a short length scale (of the order of the muffin-tin radius) that Co \textit{is} effectively Co$^{2+}$ \cite{Wadati2010,Levy2012,Berlijn2012}. This, of course, would be fully consistent with the present observations.\\[0.25mm]
\indent In conclusion, our NEXAFS data provide direct spectroscopic evidence that not only the Fe but also the Co and the As valencies remain completely unaffected across the entire doping range. The Fe and Co $L_{2,3}$ spectra are best described assuming a tetrahedral (Fe,Co)As$_4$ coordination. This finding underlines a prominent role of the hybridization between (Fe,Co) 3$d_{xy}$, $d_{xz}$, $d_{yz}$ orbitals and As $4s/4p$ states for the Fermi surface in $A$(Fe$_{1-x}$Co$_x$)$_2$As$_2$. According to our experiments in conjunction with the multiplet calculations, Fe as well as Co ions retain a valency of $+2$\@ which casts serious doubt on the widely assumed electron-doping effect induced by Co. The present data are fully consistent with the calculations of Ref.~\cite{Wadati2010,Levy2012,Berlijn2012,Liu2012} where the effect of Co substitution is described in terms of a (topological \cite{Wadati2010,Levy2012}) change of the Fermi surface (probably induced by a Lifshitz transition) which destabilizes the magnetism in favor of superconductivity. In these calculations, Co is identified either as \textit{being} isovalent to Fe \cite{Wadati2010,Levy2012,Liu2012}, or as being screened in such a way as to \textit{appear} isovalent to Fe \cite{Berlijn2012}.\\[0.25mm]
\indent The current investigation indicates a more pronounced covalent character for the (Fe,Co)-As bond with increasing Co content. Since the corresponding bond length, i.~e., the distance of the As atoms to the (Fe,Co) position, is changed by application of external pressure \cite{Alireza2009,Yamazaki2010,Drotziger2010} or upon chemical pressure due to the isovalent substitution of As by P \cite{Kasahara2010} or Fe by Ru \cite{Rullier2010}, the covalency of the (Fe,Co)-As bond seems to be a key parameter for the interplay between magnetism and superconductivity in $A$(Fe$_{1-x}$Co$_x$)$_2$As$_2$ and for the topology of the Fermi surface as well.
The strongly covalent character of the (Fe,Co)-As bond is also consistent with Refs.~\cite{Kasahara2010,Rullier2010} where it was suggested that even the mobility of charge carriers can be significantly enhanced upon the isovalent replacement of As by P or Fe by Ru.

\begin{acknowledgments}
We are indebted to D. Kronmüller, B. Scheerer, P. Adelmann, A. Assmann, and S. Uebe for their excellent technical support and for fruitful discussions. We greatly appreciate stimulating discussions with W. Ku and G. A. Sawatzky. We gratefully acknowledge the Synchrotron Light Source ANKA Karlsruhe for the provision of beamtime. Part of this work was supported by the German Science Foundation (DFG) in the framework of the Priority Program SPP1458.

\end{acknowledgments}

\bibliographystyle{apsrev}
\bibliography{pnictide2}

\begin{thebibliography}{47}
\bibitem[{foo(2011{\natexlab{c}})}]{footnote2}
\bibinfo{journal}{It should be pointed out that for Co$^{3+}$ HS in tetrahedral
  coordination, parameters in the calculations can be chosen so that the peak
  around 781~eV occurs closer to the experimentally observed position. However,
  the concomitant increase of spectral weight for the satellite structure at
  776 eV and for the shoulder on the low-energy side of the $L_2$ edge,
  together with the reduced width of the main peak at the $L_3$ edge, obviously
  disagree with the measured data. In all cases, the simulations shown are
  those with the best agreement with the experimental data.}
  (\bibinfo{year}{2011}{\natexlab{c}}).
\end{thebibliography}

\end{document}